\documentclass[aps,pra,showpacs,twocolumn,amsmath,amssymb,superscriptaddress,nofootinbib]{revtex4}
\usepackage{graphicx}
\usepackage{bm}
\usepackage{amssymb}
\usepackage{amsmath}
\usepackage{latexsym}
\usepackage{color}
\newcommand{\ind}[1]{_{#1}}    
\newcommand{\indrm}[1]{_{\mathrm {#1}}}    

\newcommand{\fwhms}{$\sigma_{\ind{\Delta \theta}}$}
\newcommand{\fwhmsa}{$\left< \sigma_{\ind{\Delta \theta}}\right>$}
\newcommand{\fwhmss}{$\sigma_{\ind{\Delta \theta}}^{*}$}
\newcommand{\fwhmssa}{$\left< \sigma_{\ind{\Delta \theta}}^{*}\right>$}
\newcommand{\fwhmsn}{$\Sigma_{\ind{\Delta \theta}}$}
\newcommand{\fwhmsan}{$\left< \Sigma_{\ind{\Delta \theta}}\right>$}
\newcommand{\fwhmssn}{$\Sigma_{\ind{\Delta \theta}}^{*}$}
\newcommand{\fwhmssan}{$\left< \Sigma_{\ind{\Delta \theta}}^{*}\right>$}
\newcommand{\coms}{$\sigma_{\ind{\theta}}$}
\newcommand{\comss}{$\sigma_{\ind{\theta}}^{*}$}
\newcommand{\comsa}{$\left< \sigma_{\ind{\theta}} \right>$}
\newcommand{\comssa}{$\left< \sigma_{\ind{\theta}}^{*} \right>$}
\definecolor{ncolor}{rgb}{0.,0.,0.}

\begin{document}  

\title{Small Bragg-plane slope errors revealed in synthetic diamond crystals}

\author{Paresh Pradhan}
\author{Michael Wojcik}
\author{Xianrong Huang}
\author{Elina Kasman}
\author{Lahsen Assoufid}
\author{Jayson Anton}
\author{Deming Shu}
\affiliation{Advanced Photon Source, Argonne National Laboratory,  Argonne, Illinois 60439, USA}
\author{Sergey Terentyev}
\author{Vladimir Blank}
\affiliation{Technological Institute for Superhard and Novel Carbon  Materials, 142190~Troitsk, Russian Federation}
\author{Kwang-Je Kim}
\author{Yuri Shvyd'ko} \email{shvydko@anl.gov}
\affiliation{Advanced Photon Source, Argonne National Laboratory,  Argonne, Illinois 60439, USA}

\begin{abstract} 
   Wavefront-preserving x-ray diamond crystal optics are essential for
   numerous applications in x-ray science.  Perfect crystals with flat
   Bragg planes are a prerequisite for wavefront preservation in Bragg
   diffraction. However, this condition is difficult to realize in
   practice because of inevitable crystal imperfections. Here we use
   x-ray rocking curve imaging to study the smallest achievable
   Bragg-plane slope errors in the best presently available synthetic
   diamond crystals and how they compare with those of perfect silicon
   crystals. We show that the smallest specific slope errors in the
   best diamond crystals (both freestanding or strain-free mounted)
   are about 0.15-0.2~$\mu$rad/mm$^2$. These errors are only a factor
   of two larger than the 0.05-0.1~$\mu$rad/mm$^2$ specific slope
   errors we measure in perfect silicon crystals. High-temperature
   annealing at 1450$^{\circ}$C of almost flawless diamond crystals
   reduces the slope errors very close to those of silicon. Further
   investigations are required to establish the wavefront-preservation
   properties of these crystals.
\end{abstract}

\maketitle

\section{Introduction}

Diamond features a unique combination of outstanding physical
properties perfect for numerous x-ray crystal optics applications
where traditional materials such as silicon fail to perform.  Diamond
is a material of choice in applications requiring improved
transparency to x-rays, highest x-ray Bragg reflectivity, thermal
conductivity, mechanical stiffness, and resilience to radiation
damage.  Diamond optics are essential for tailoring x-rays to the most
challenging needs of x-ray research. Diamond optics are becoming vital
for generation of fully coherent hard x-rays by seeded x-ray
free-electron lasers [see recent review paper \cite{SBT17} for details
  and references].

Progress in fabrication of synthetic high-quality diamond crystals has
been substantial in the last two decades.  Crystals with defect-free
areas of $\simeq 4\times4$~mm$^2$ and more grown by a temperature
gradient method under high pressure and high temperature (HPHT) are
now state of the art \cite{BCC09,PDK11,SSB11,ST12,SAB16}.  However,
the perfection of diamond crystals is typically not as high as of
silicon crystals, which are standard in x-ray crystal optics
applications. In particular, the wavefront-preservation properties,
critical for many applications, suffer from insufficient crystal
quality.

Perfect crystals with flat Bragg planes are a prerequisite for
wavefront preservation in Bragg diffraction. But nothing is perfect.
How flat can Bragg crystal planes be in the best available diamond
crystals?  What are the smallest achievable Bragg-plane slope errors
in the best presently available synthetic diamond crystals? How
do these compare to those in perfect silicon crystals?  These questions
are addressed in the present paper.

In the studies presented in this paper, Bragg-plane slope errors are
measured using x-ray Bragg diffraction {\color{ncolor} rocking curve
  imaging (RCI), also known as sequential topography \cite{LBH99}.}
This technique is applied to best available diamond crystals featuring
relatively large areas ($\simeq 4\times 4$~mm$^2$) almost free of
dislocations, stacking faults, inclusions, and other defects
detectable by white beam x-ray topography \cite{Tuomi74,BoTa98}, which
is used to prescreen the diamond crystals. The Bragg-plane slope
errors in diamond crystals are compared to those in highest quality
reference silicon crystals.

We show that the smallest specific slope errors in the best diamond
crystals are about \comss ~$\simeq$ 0.15-0.2~$\mu$rad/mm$^2$, which are
only a factor of two larger than the $\lesssim 0.1~\mu$rad/mm$^2$
slope errors we measure in reference silicon crystals. Such small
slope errors are achieved not only in freestanding diamond crystals
but also in crystals firmly mounted in crystal holders, provided the
crystals are designed and machined with special strain-relief
features.  High-temperature annealing at 1450$^{\circ}$C of the best
diamond crystals further reduces Bragg-plane slope errors, such that hey approach
those of silicon.

RCI data also provides access  to the specific dispersion \fwhms\ of
the rocking curve widths $\Delta \theta$. Normalized to the Bragg
reflection width $\Delta \theta$, it is a measure of the deviation
from the largest Bragg reflectivity achievable by perfect
crystals. The best diamond crystals feature normalized specific dispersion
values \fwhmssn $\simeq$ 0.01-0.013/mm$^2$ vs. $\simeq$
0.003-0.005/mm$^2$ in silicon. These data indicates that the local
reflectivity values in the best diamond crystals are reduced by not
more than 1\% to 1.3\% from the maximum values, in agreement with
previous Bragg reflectivity studies in diamond \cite{SSB11}.

Further investigations are required to establish the
wavefront-preservation properties of the best available diamond
crystals.

The paper is organized as follows. In Section~\ref{Si} we provide
results of the RCI studies in a reference silicon crystal.  Results of
studies in selected freestanding diamond crystals and comparison with
the reference silicon crystal are presented in
Section~\ref{Diamond-free}. Design, fabrication, and RCI studies of
diamond crystals with strain relief features mounted in crystal
holders are discussed in Section~\ref{Diamond-mounted}. Effects of
high-temperature heat treatment on the Bragg-plane slope errors in diamond
crystals are discussed in Section~\ref{Diamond-annealing}.  We refer
to Appendix~\ref{method} for details on the RCI technique and to
Appendix~\ref{moderate} for the methods of mitigating the impact of
the beamline wavefront distortions on the actual values of the  Bragg-plane slope
errors.  Appendix~\ref{HTHV} provides details on
high-temperature annealing.

\section{Bragg-plane slope errors in reference silicon}
\label{Si}

Prior to studying Bragg-plane slope errors in diamond crystal, we used
the same RCI technique and setup (introduced in Appendix~\ref{method})
to measure RCI maps and the relevant crystal parameters in a specially
prepared reference silicon crystal. These measurements were performed to
establish a reference for the diamond crystals and to benchmark the
ultimate performance of the RCI setup used in the later studies.

The reference crystal was manufactured from the highest quality
{\color{ncolor} high-resistivity} single-crystal silicon, with the (531) crystal planes
parallel to the surface. The 531 Bragg reflection is used to match the!
531 Bragg reflection from the conditioning crystal. The crystal was
made relatively large (15$\times$15$\times$15~mm$^2$). Its lower
part was fixed in a crystal holder in a manner that did not create strain in the
upper part, which was exposed to x-rays.

\begin{figure}
  \includegraphics[width=0.5\textwidth]{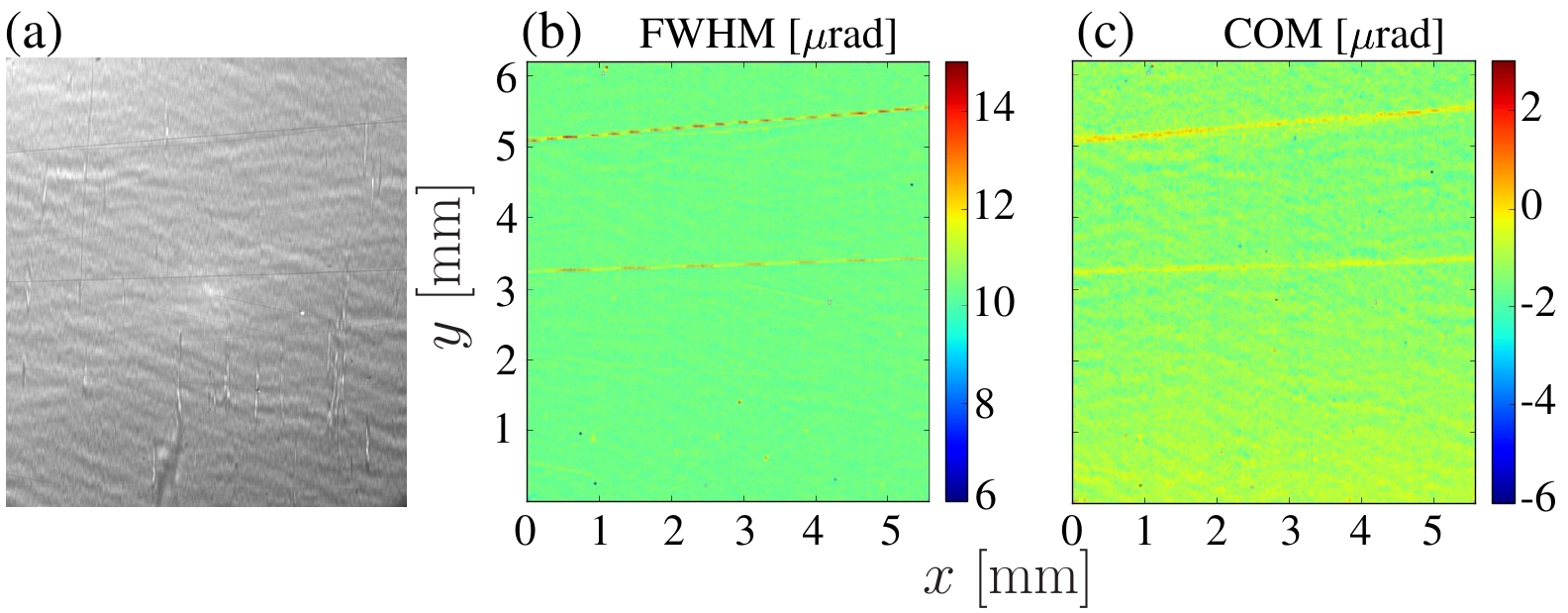}
  \caption{X-ray image and rocking curve maps of a reference silicon crystal
  in the 531 Bragg reflection. (a) X-ray image at the peak of the
  crystal-integrated rocking curve. (b) Color map of the Bragg
  reflection angular widths (FWHM) and (c) color map of the center of
  mass (COM) of the Bragg reflection angular dependences. All color
  maps presented in the paper are calculated with binning number
  $N=10$.}
\label{fig001a}
\end{figure}

Figure~\ref{fig001a}(a) presents an x-ray image of our reference silicon
crystal in the 531 Bragg reflection recorded at the crystal-integrated
Bragg reflection rocking curve maximum [displayed in
  Fig.~\ref{fig022}(a) of Appendix~\ref{method}]. Figure~\ref{fig001a}(b) shows a color map of
the Bragg reflection angular widths $\Delta \theta$ (full width at
half maximum, FWHM), while Fig.~\ref{fig001a}(c) shows a color map of
the Bragg reflection peak relative angular positions $\theta$,
evaluated as a center of mass (COM) of the rocking curves.  We note
that the rippled background and the two straight lines in
Figs.~\ref{fig001a}(a)-(c) are artifacts caused by beamline optical
components (see Appendix~\ref{moderate} for more details).

The RCI data provide access to numerous crystal parameters that are
calculated by the RCI data evaluation code.  The parameters used in
these studies and their definitions are summarized Table~\ref{tab1}.

\begin{figure}
        \includegraphics[width=0.5\textwidth]{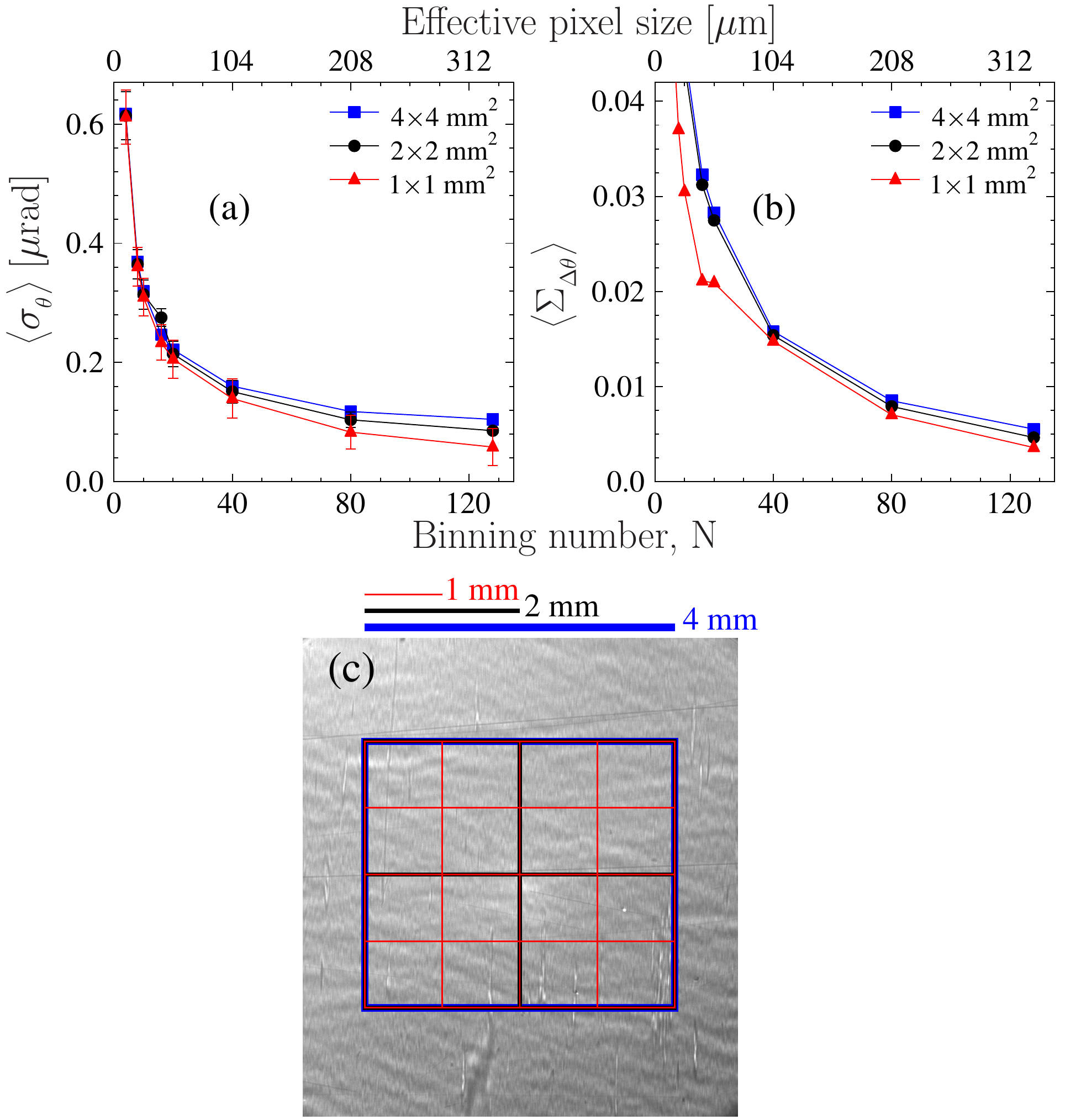}
  \caption{Averaged Bragg-plane slope errors \comsa\ (a) and
      normalized dispersions \fwhmsan\ of the Bragg reflection widths
      (b) for the reference Si crystal in the 531 Bragg reflection
      measured and calculated as a function of the area detector
      binning number $N$. The averaging is performed in one of three
      ways over sixteen equal-size $1\times 1$~mm$^2$ crystal's ROIs,
      over four $2\times 2$~mm$^2$ ROIs, or just calculated in one
      $4\times 4$~mm$^2$ ROI, as indicated by red, black, and blue
      lines, respectively, in (a) and (b) and on the x-ray Bragg
      diffraction image in (c). }
\label{fig002a}
\end{figure}

Of these parameters, Bragg-plane slope error \coms\ is one of the most
important in the present studies. Figure~\ref{fig002a}(a) shows
\comsa\ values in the reference silicon crystal as a function of the
binning number calculated in different regions of interest (ROI)
indicated by appropriate colors in Fig.~\ref{fig002a}(c).  The error
bars represent variation of the \coms\ values being averaged.

\begin{table}
    \begin{tabular}{p{2.7cm} p{5.7 cm}}
  \hline Notation & Definition\\ \hline

$\theta$ & Relative angular position of the Bragg reflection peak
(center of mass, COM) measured at a particular location on the crystal
by an area detector pixel.\\[1mm]

\coms\ & Bragg-plane slope error calculated as dispersion of the
$\theta$ values within a selected region of interest (ROI) on the crystal.\\[1mm]

\comsa\ & \coms\ values averaged over multiple equal-size ROIs.\\[1mm]

\comss, \comssa\  & Specific \coms\ or \comsa\  values measured  over $1\times 1$~mm$^2$ ROIs.\\[1mm]

\hline

$\Delta \theta$ & Angular width (full width at half maximum, FWHM) of
  the Bragg reflection peak measured locally on the crystal by an area
  detector pixel.\\[1mm]


\fwhms\ & Dispersion of the $\Delta \theta$ values  within  an ROI.\\[1mm]

\fwhmsa\ & \fwhms\  values averaged over several equal-size ROIs.\\[1mm]

\fwhmss, \fwhmssa\ & Specific \fwhms\ or \fwhmsa\ values measured  over $1\times 1$~mm$^2$ ROIs.\\[1mm]

\fwhmsn =2.35\fwhms/$\Delta \theta$, \fwhmssn =2.35\fwhmss/$\Delta \theta$, etc  &  Normalized \fwhms, \fwhmss, \fwhmsa\  or \fwhmssa\ values.\\[1mm]

\hline
$N$ & Data binning number. It is related to a method of adding
(binning) the signal from adjacent $N\times N$ area detector pixels
together to achieve better signal-to-noise ratio or to minimize the
effects of small observation errors at a cost of resolution.\\[0mm]

\hline

\end{tabular}
\caption{Notations and definitions of the characteristic crystal
  values measured and evaluated with the x-ray rocking curve imaging
  (RCI) technique.}
\label{tab1}
\end{table}

The binning procedure moderates the impact of imperfections in the
beamline optics, as discussed in more detail in
Appendix~\ref{moderate}. Remarkably, there is no significant change in
the \comsa\ values measured in ROIs of different size, indicating a
homogeneous quality of the reference silicon crystal and fairly flat
crystal planes over relatively large crystal areas. The average
specific Bragg-plane slope errors are \comssa $\simeq 0.1-0.05
~\mu$rad/mm$^2$. These numbers may not necessarily represent the true
value of the Bragg-plane slope errors in silicon crystals. They might
be even smaller. These numbers rather represent the resolution of our
setup, limited by wavefront distortions in the beamline x-ray optical
components.

RCI data also provide access to the dispersion \fwhms\ of the rocking
curve widths $\Delta \theta$. Normalized to the Bragg reflection width
$\Delta \theta$, it is a measure of the deviation from the largest
Bragg reflectivity value for the given reflection. This can be easily
understood, because the product of the Bragg reflection width and the
reflectivity is an invariant value in the first approximation.
Figure~\ref{fig002a}(b) shows a plot of the normalized and averaged
Bragg width dispersion \fwhmsan\ values (see Table~\ref{tab1} for the
definition) in the reference silicon crystal as a function of the
binning number calculated in ROIs of different size.  Similar to the
case of \comsa\ values in Fig.~\ref{fig002a}(a), there is no
significant change in the \fwhmsn\ values with the size of ROI. This
is another indication of a very high and homogeneous quality of the
reference silicon crystal. The specific average normalized Bragg width
dispersion in silicon is \fwhmssan $\simeq$ 0.003-0.005/mm$^2$. This
is a small value, which indicates that the maximum Bragg reflectivity
maybe reduced by less than  0.5\% due to crystal strain.
Similar to the 
case of \comssa , the small value of \fwhmssan\ we  measure in
silicon may represent the resolution limit of the setup rather  than
the real value for single crystal silicon, which maybe even smaller.

\begin{figure}
    \includegraphics[width=0.5\textwidth]{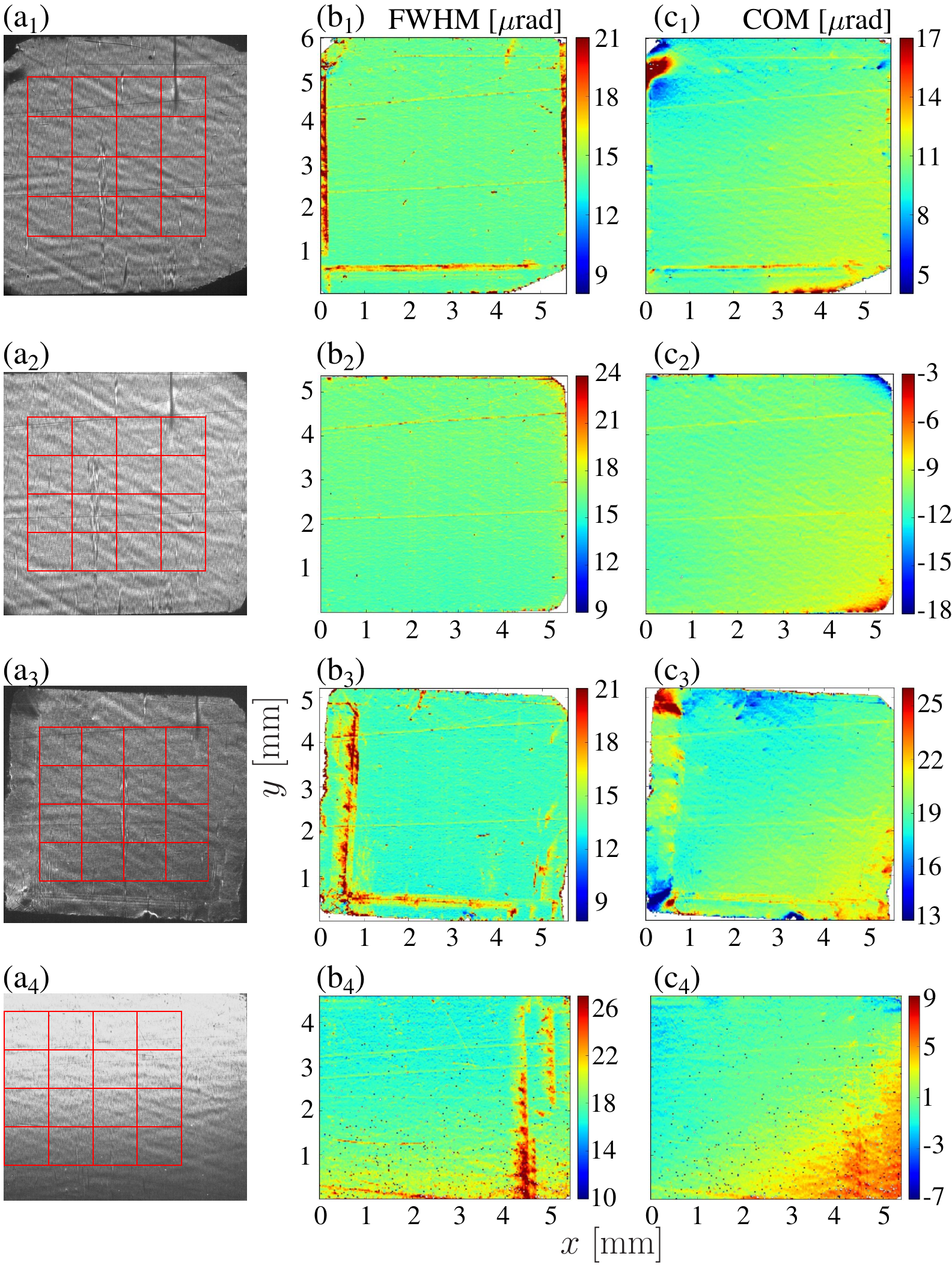}
    \caption{X-ray 400 Bragg reflection images and RCI maps for four
    selected type IIa HPHT diamond crystal plates in the (100)
    orientation. Columns are as in Fig. 1: (a) x-ray image at peak of
    crystal-integrated rocking curve, (b) Bragg reflection angular
    width, and (c) center of mass of the Bragg reflection angular
    dependences. Rows correspond to crystals: (1) VB4, (2) VB5, (3)
    VB6, and (4) D3.  The red grids in column (a) indicate ROIs,
    similar to Fig.~\ref{fig002a}(c). }
\label{fig001}
\end{figure}

\section{Bragg-plane slope errors in diamond crystals}
\label{Diamond-free}

The studies of the Bragg-plane slope errors in diamond crystals were
performed on samples selected using white-beam x-ray topography. All
crystals are of type IIa grown by HPHT technique, cut and polished to
plates in the (100) orientation \cite{PDK11}.  Two of the available
crystals (termed VB4 and VB5) feature large areas ($\simeq 4\times
4$~mm$^2$) free of dislocations, stacking faults, inclusion, and other
defects detectable by white beam x-ray topography. Crystal VB4 was
used previously in diamond Bragg reflectivity studies
\cite{SSB11}. X-ray Bragg diffraction images and RCI maps of these two
crystals are shown in the upper two rows of Fig.~\ref{fig001}. The RCI
maps are homogeneous in the central region, revealing in particular a
Bragg reflection width of $\Delta\theta=14.8~\mu$m close to the
theoretical value.  Another crystal (termed VB6) features a few weak
dislocation lines in the central part.  Stacking faults at the edges
result is a propagating strain fields as well as areas with enlarged
rocking curve widths clearly seen on the RCI maps in row 3 of
Fig.~\ref{fig001}.  X-ray white-beam topography reveals more defects
and propagating strain fields in a crystal labeled D3.  The relevant
RCI data in row 4 reveal crystal quality inferior to that of crystals
VB4, VB5, and even VB6. All crystal plates are rather thick: crystal
VB4 is 1~mm, while the others are about 0.5 to 0.6~mm thick.

To avoid any externally induced strain, which could be caused for
example by crystal mounting, the crystals lie free in a flat 1-mm-deep
indentation machined in an aluminum block fastened to the stage used
to perform angular scans. The indentation holding the diamond crystal
in it was covered with a thin plastic foil to minimize the effect of air
circulation on the angular stability of the crystal.

\begin{figure}
      \includegraphics[width=0.5\textwidth]{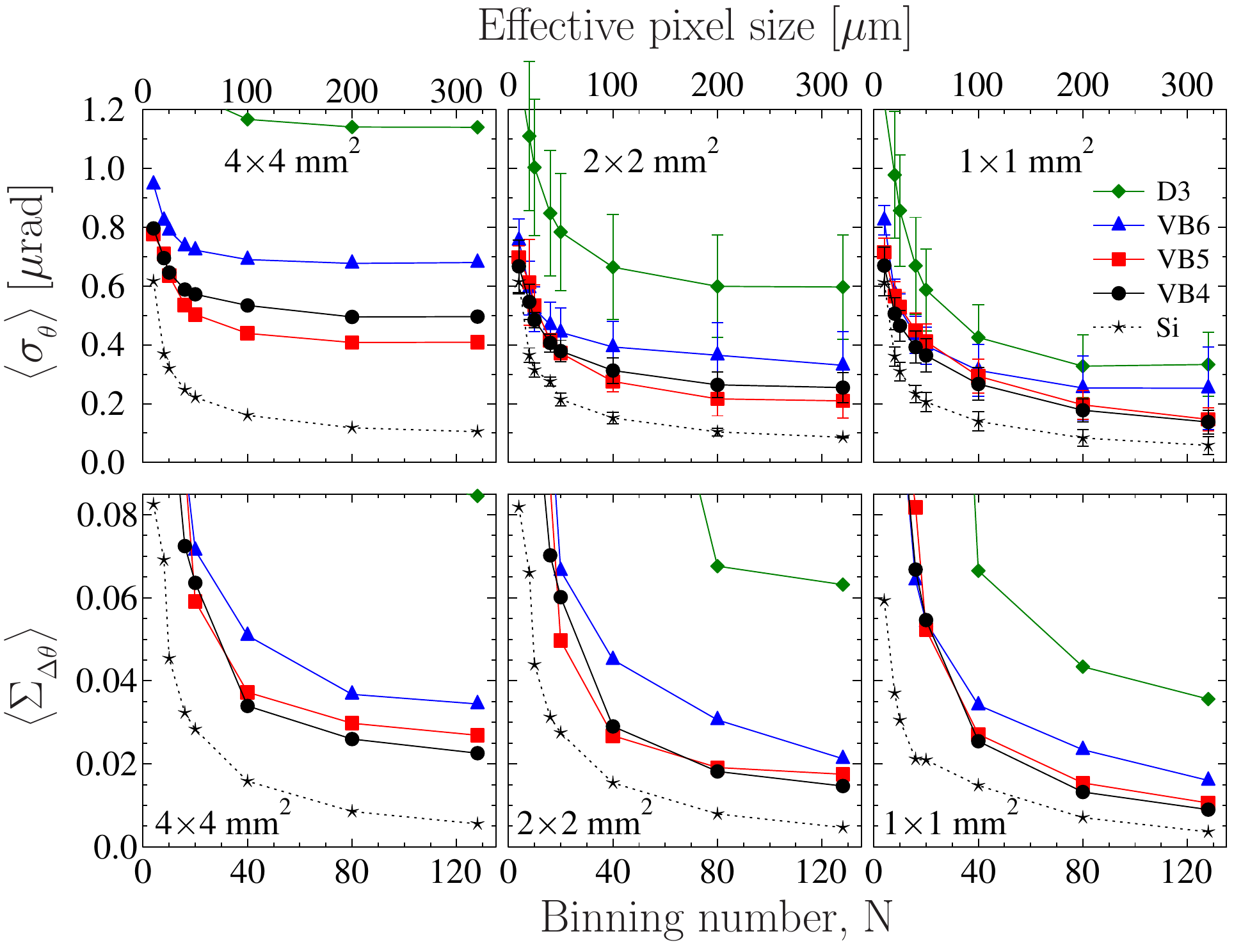}    
  \caption{Averaged Bragg-plane slope errors \comsa\ (upper row)
    together with averaged and normalized Bragg-reflection-widths
    dispersions \fwhmsan\ (lower row) in selected diamond crystals
    measured in the 400 Bragg reflection and calculated from the RCI
    data as a function of the area detector binning number $N$. The
    data evaluation and averaging is performed in selected ROIs
    indicated by red grids in Fig.~\ref{fig001} in a similar way as in
    Fig.~\ref{fig002a}. }
\label{fig002}
\end{figure}

The upper row of Fig.~\ref{fig002} showns the averaged Bragg-plane
slope errors \comsa\ measured and calculated in ROIs of different size
for the four selected freestanding crystals. Data for the reference
silicon crystal are also presented there for reference. The best
crystal regions with the lowest density of defects are selected for
this purpose, as indicated by red grids in
Figs.~\ref{fig001}(a$_{\ind{1}}$)-(a$_{\ind{4}}$).  The smallest slope
errors are observed as expected in crystals VB4 and VB5, in which the
defects appear only on the crystal rims outside the selected ROIs.
Unlike in silicon, slope errors in diamond crystals, even in the best
crystals (VB4 and VB5), change substantially with the size of the
ROI. This result indicates that the diamond crystal quality is less
homogeneous than that of silicon.

Nevertheless, the averaged specific slope errors \comssa\ in the two
best diamond crystals VB4 and VB5 (practically flawless in the central
$4\times 4$~mm$^2$ ROI) feature values \comssa
$\simeq$~0.15-0.2~$\mu$rad/mm$^2$, only a factor of two larger than
those in silicon. The overlapping error bars indicate that in some
ROIs the specific errors in diamond and silicon are even comparable.

Similarly, the graphs in the lower row of Fig.~\ref{fig002} present
plots of the averaged and normalized Bragg-reflection-width
dispersions \fwhmsan\ in the selected diamond crystals measured and
calculated from the RCI data as a function of the area detector
binning number $N$. The \fwhmsan\ values calculated in ROIs of various
sizes are quite different, thus revealing again, in agreement with the
 \comsa\ values, more inhomogeneities in the diamond crystals
than in the reference silicon crystal. Nevertheless, the specific
values presented in the 1$\times$1-mm$^2$ graph can be small,
especially for the highest quality crystals (VB4 and VB5), approaching
\fwhmssan ~$\simeq$~0.01-0.013/mm$^2$, only a factor of two larger
than the relevant reference silicon values.

This data indicates that the local reflectivity values in the two best
diamond crystals are reduced (possibly by residual crystal strain) by
no more than 1\% to 1.3\% from the maximum possible value. This result
is in agreement with the direct absolute reflectivity measurements
previously performed on crystal VB4  \cite{SSB11}. The peak reflectivity
measured with an x-ray beam having a  1$\times$1~mm$^2$ in cross-section and
averaged over the central crystal area of 1$\times$4~mm$^2$ was
99.1\%$\pm$0.4\%, which was close to the theoretical value of 99.7\%. We note
that the  specific \fwhmssan\ values presented here are evaluated on a
larger crystal area of 4$\times 4$~mm$^2$.

\section{Impact of diamond crystal clamping}
\label{Diamond-mounted}

The data presented in the previous section were obtained on
freestanding crystals. However, for optical components to function
properly, they must be rigidly mounted in crystal holders to ensure
angular and position stability.  Correct mounting also provides for
thermal transport to discharge the x-ray beam power absorbed by the
crystal.

Even though the selected crystal plates are 0.5 to 1-mm thick and
therefore very stiff due to the very large Young's modulus of diamond,
clamping without any precautions produce tremendous strain. A standard
approach of reducing mounting strain is to introduce strain-relief
features.

\begin{figure}
\includegraphics[width=0.43\textwidth]{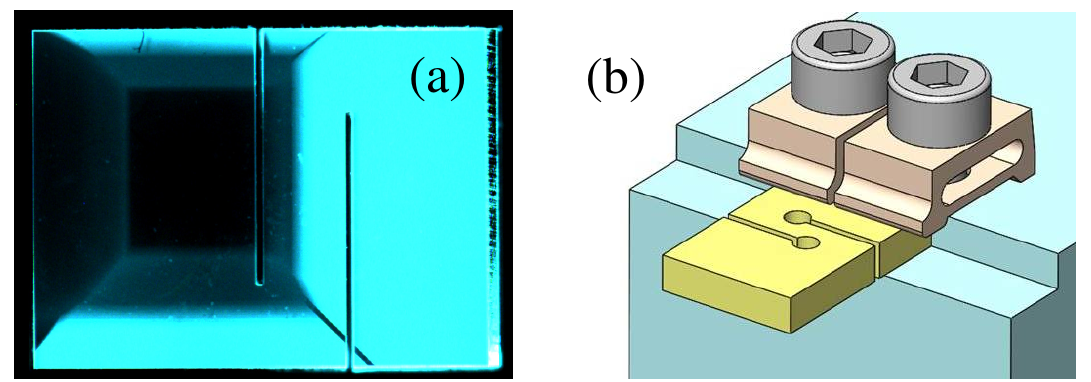}  
\caption{Specially designed and manufactured type IIa HPHT diamond
  crystal plate in the (100) orientation furnished with strain-relief
  cuts. (a) UV-excited luminescence image.  (b) Schematic  of
  the crystal with strain relief features clamped in a crystal holder. }
\label{st2-1ab}
\end{figure}

A high-quality type IIa HPHT 440-$\mu$m-thick diamond crystal plate in
the (100) orientation was selected, featuring a small amount of
defects, mostly at the crystal rim.  It was cut to a rectangular
$5.4\times4.5$~mm$^2$ plate and furnished with strain relief cuts, as
seen on a UV-excited luminescence image in Fig.~\ref{st2-1ab}. The
darkest zone in the UV image corresponds to the (100) growth sector
with lowest nitrogen content.  The strain-relief cuts are the two
vertical parallel dark lines. The cuts were introduced to prevent
propagation of strain into the working area (on the left of left cut)
provided the crystal is clamped rigidly on the right of the right
cut. The cuts were made with YAG:Nd laser pulses in the 2nd harmonic
with 100-ns duration, 1.7-mJ/pulse energy, spot size 20- to 25-$\mu$m,
and 5-kHz repetition rate. The width of the cuts is $\simeq 50~\mu$m,
made in two passes with a 25-$\mu$m lateral shift.  Finite-element
analysis shows that adding holes at the end of the cuts, as shown in
Fig.~\ref{st2-1ab}(b), produces better strain relief; however, such
holes were not implemented for this particular sample.

\begin{figure}
  \includegraphics[width=0.5\textwidth]{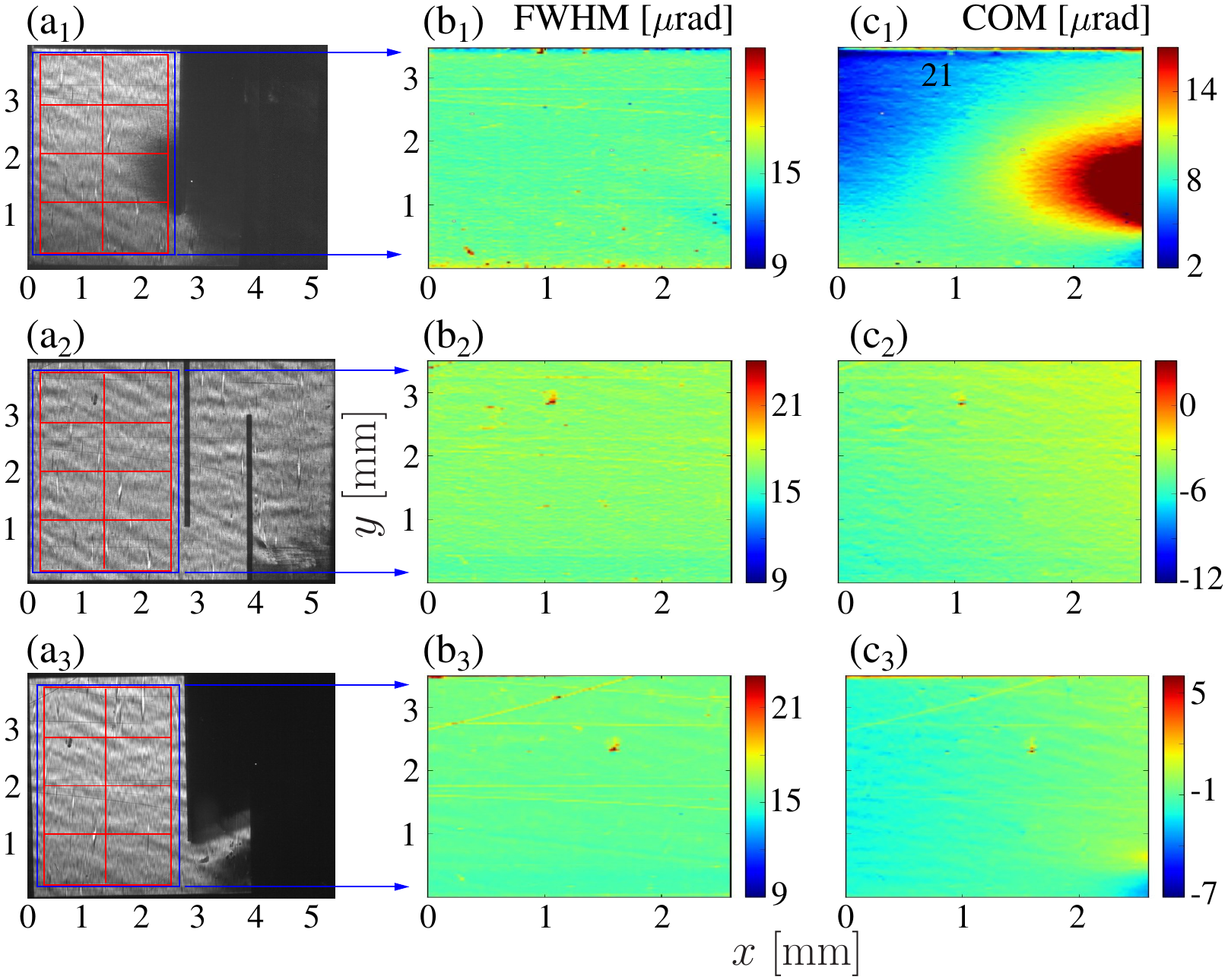}  
\caption{X-ray 400 Bragg reflection images and RCI maps for mounted type
  IIa HPHT diamond crystal with strain-relief cuts. Columns are as in
  Fig.~\ref{fig001}: (a) X-ray 400 Bragg reflection image at peak of crystal-integrated rocking curve,
  (b) Bragg reflection angular width, and (c) center of mass of the
  Bragg reflection angular dependences in the ROIs indicated by the
  red grids in column (a). Rows correspond to the following
  conditions: (1) after laser cutting, (2) after annealing in air at
  630$^{\circ}$C for 3 hours, (3) after clamping the right part in the
  crystal holder.}
\label{st2-1}
\end{figure}

Figure~\ref{st2-1}(a$_{\ind{1}}$) shows an x-ray 400 Bragg diffraction
image at the top of the crystal integrated rocking curve.  The image
reveals that the laser cutting induces a very large strain: only the
working area reflects x-rays and can be imaged. The rest of the
crystal is out of reflection because of the cutting-induced strain.  The FWHM
map in the working area shown in Fig.~\ref{st2-1}(b$_{\ind{1}}$) is
very homogeneous, revealing almost theoretical Bragg reflection width,
which proves almost defect-free crystal quality.  However, the COM map
presented in Fig.~\ref{st2-1}(c$_{\ind{1}}$) reveals very large
strain, in agreement with Fig.~\ref{st2-1}(a$_{\ind{1}}$).

Is it possible to eliminate strain induced by laser cutting?
In our previous studies \cite{KVT16} we found that annealing diamond
crystals in air at a temperature of 630--650$^{\circ}$C for 3 hours may
substantially reduce strain induced in the process of laser cutting or
ablation. The strain is caused by the graphitization of the machined
surfaces.  The annealing temperature is chosen such that all residuals
of graphite and other carbon compounds are burned in air, while
keeping diamond intact\footnote{Increasing annealing temperature or
  time may result in etching of the crystal surface.}. We will refer to
this procedure in the following as
medium-temperature in-air annealing (MTA).

Indeed, such annealing practically erases the cutting-induced strain
as the x-ray Bragg diffraction image in
Fig.~\ref{st2-1}(a$_{\ind{2}}$) and RCI maps in
Figs.~\ref{st2-1}(b$_{\ind{2}}$)-(c$_{\ind{2}}$) evidence.  These
measurements were performed on a free-standing crystal in the
configuration previously described.

X-ray Bragg diffraction images and RCI maps in
Figs.~\ref{st2-1}(a$_{\ind{3}}$)-(c$_{\ind{3}}$) show what happens to
the crystal if it is rigidly clamped, as schematically presented  in
Fig.~\ref{st2-1ab}(b). At a first glance the image in
Fig.~\ref{st2-1}(a$_{\ind{3}}$) resembles the case of
Fig.~\ref{st2-1}(a$_{\ind{1}}$): only the working area can be imaged,
while the rest is heavily strained and is out of reflection. In
reality, the new situation is completely different.  The COM map of
the working area of the clamped crystal (on the left of the left cut)
in Fig.~\ref{st2-1}(c$_{\ind{3}}$) looks very similar to the COM map
of the freestanding annealed crystal in
Fig.~\ref{st2-1}(c$_{\ind{2}}$). This demonstrates that clamping of a
crystal furnished with strain-relief cuts does not produce strain in
the working area.

\begin{figure}
      \includegraphics[width=0.5\textwidth]{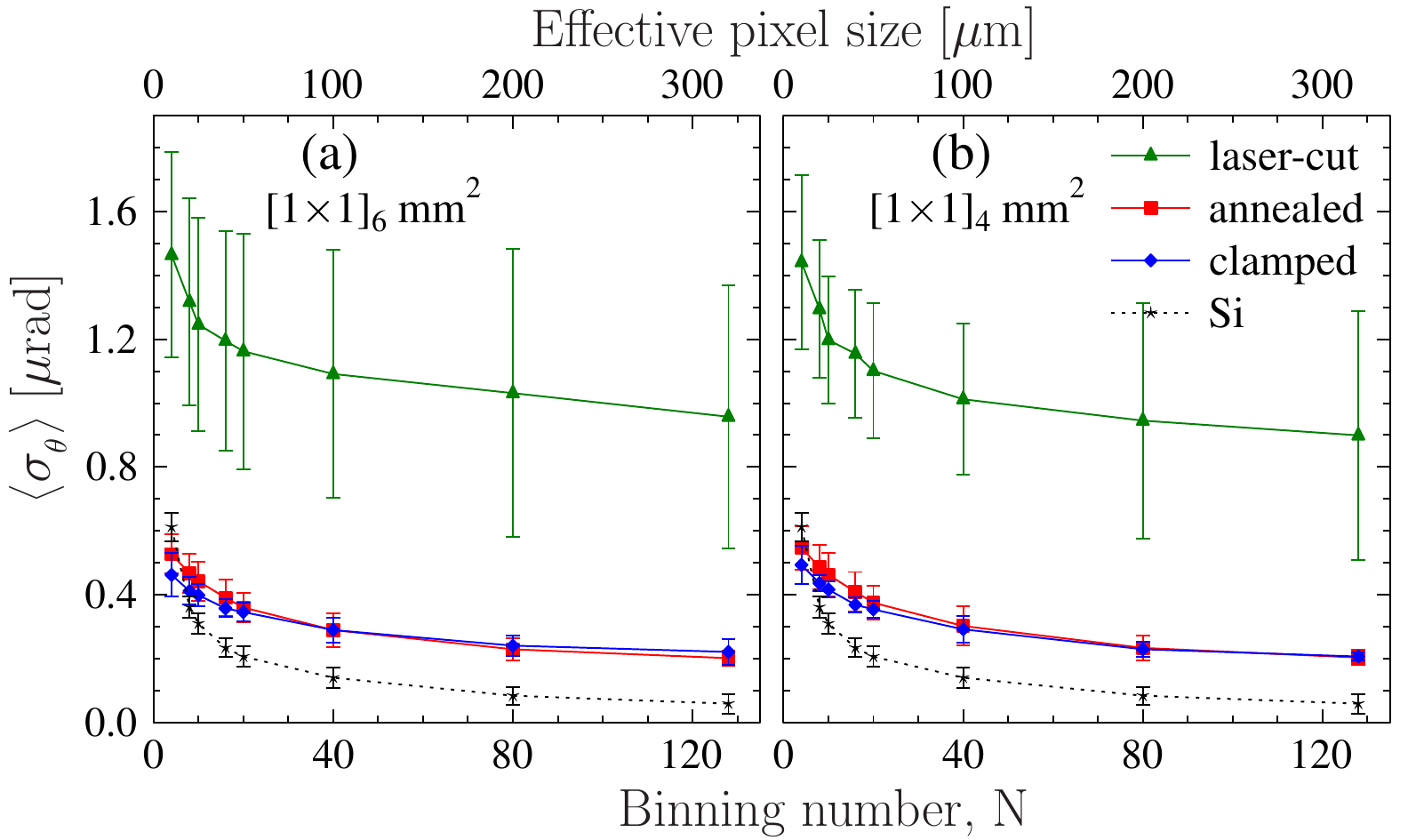}  
\caption{Averaged specific Bragg-plane slope errors \comssa\ in the
  diamond crystal with strain-relief cuts after laser cutting (green),
  after MTA annealing (red), and after clamping (blue), evaluated as a
  function of the area detector binning number $N$ and compared to the
  reference silicon crystal values. The averaging is performed over
  either (a) six or (b) four equal-size $1\times 1$~mm$^2$ ROIs.}
\label{st2-2}
\end{figure}

Averaged specific Bragg-plane slope error values \comssa\ presented in
Fig.~\ref{st2-2} support this statement.  Both graphs show
\comssa\ dependences on the binning number for the crystal after laser
machining (green lines and markers), after MTA annealing (red), and
clamped (blue).  The averaging is performed either over six equal-size
$1\times 1$~mm$^2$ ROIs, shown in Fig.\ref{st2-2}(a) or over four such
ROIs, shown in Fig.\ref{st2-2}(b). The ROIs are indicated by red grids
in Figs.~~\ref{st2-1}(a$_{\ind{1}}$), (b$_{\ind{1}}$), and
(c$_{\ind{1}}$), respectively.  These data demonstrate that, first,
annealing indeed helps to reduce substantially the slope errors to the
values \comssa ~$\simeq 0.2~\mu$rad/mm$^2$, which are very close to
those observed for the best freestanding diamond crystals as
documented in Fig.~\ref{fig002}~(upper right graph). Second, and most
important, the clamping does not degrade the observed slope errors in
the upper working area. Extending averaging to all eight equal-size
$1\times$ 1~mm$^2$ ROIs results in an increase of \comssa\ (data not
shown), indicating that the working area should be limited to the top
2$\times$2~mm$^2$ or at most to the 2$\times$6~mm$^2$ zone.

\section{Effect of high-temperature annealing}
\label{Diamond-annealing}

In the previous section it was shown that medium-temperature
 annealing at $\simeq$~630--650$^{\circ}$C of diamond crystals in air
helps to erase crystal strain induced by laser machining and improve
slope errors to the baseline values.

Here we study the effect of  annealing diamond crystals at
higher temperatures. Vacancies, impurity atoms such as  nitrogen, and some
other nanoscale crystal defects cannot be detected by x-ray
topographies, but they still may contribute to the Bragg-plane slope
errors.  The mobility of vacancies, impurity atoms, and other defects
increases at higher temperatures in condensed matter systems
\cite{CH96}. The expectations are that in this process the defects
may be pushed to the crystal surfaces and growth zone boundaries where they
annihilate and reduce strain.  In diamond such processes start
at about 900$^{\circ}$C, but the highest temperature should be kept
substantially lower than $\simeq$2450$^{\circ}$C, the Debye-Waller
temperature of diamond.

In our experiments, we anneal diamond crystals at 1450$^{\circ}$C for
3 hours under high-vacuum conditions ($\simeq 4\times
10^{-6}$~mbar).  We refer to this procedure  as
high-temperature high-vacuum (HTHV) annealing; 
Appendix~\ref{HTHV} gives technical details.\\

\begin{figure}
  \includegraphics[width=0.5\textwidth]{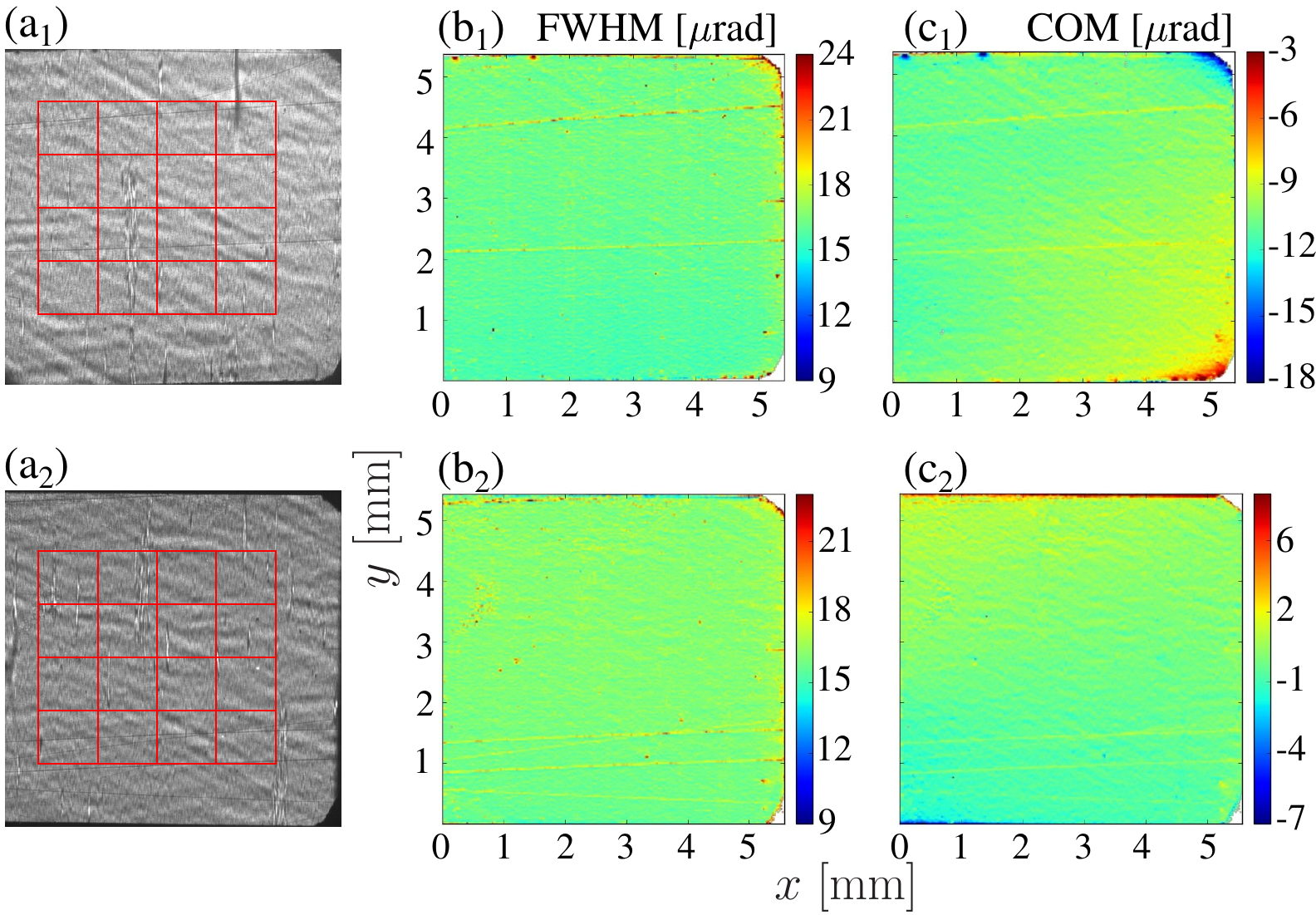}    
  \caption{X-ray 400 Bragg reflection images and RCI maps for diamond crystal VB5 after annealing. Columns are as in Fig.~\ref{fig001}: (a) x-ray image at peak of crystal-integrated rocking curve, (b) Bragg reflection angular width, and (c) center of mass of the Bragg reflection angular dependences. Rows correspond to (1) medium-temperature annealing in air (MTA) and (2) high-temperature high-vacuum annealing (HTHV).}
\label{fig013}
\end{figure}

Figure~\ref{fig013} shows x-ray 400 Bragg diffraction images and RCI maps of diamond crystal
VB5, one of the two best crystals used in these studies. The top row
shows the results after MTA but before HTHV annealing, while
the bottom row presents data after HTHV annealing. The upper row
is the same as row 2 in Fig.~\ref{fig001}. There is a clear improvement in
the homogeneity of the COM map, indicating also reduction of the
Bragg-plane slope errors \coms .

\begin{figure}
  \includegraphics[width=0.5\textwidth]{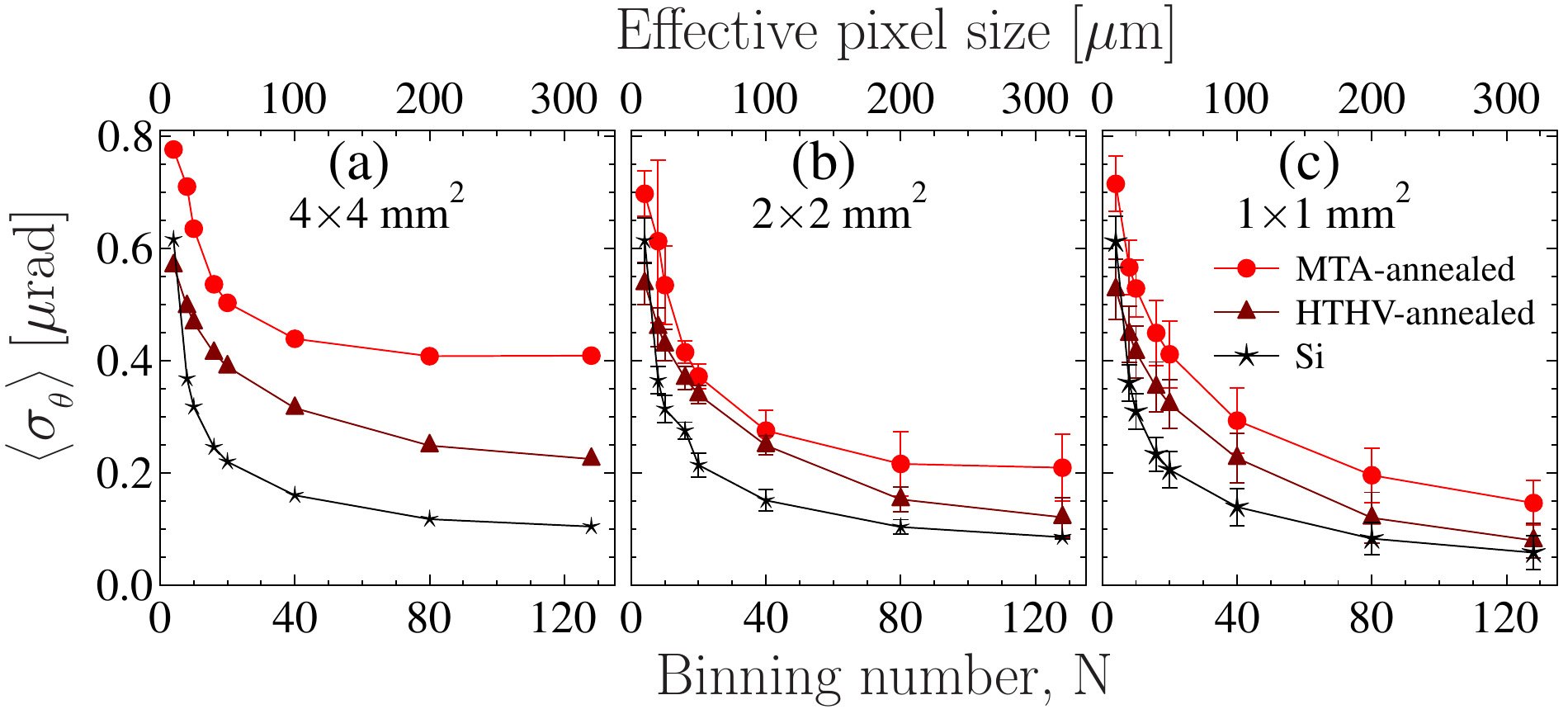}  
  \caption{Similar to Fig.~\ref{fig002}(top), here showing averaged
  Bragg-plane slope errors \comsa\ in VB5 crystal before (red markers and curves, same as in
  Fig.~\ref{fig002}) and after the HTHV annealing (brown) together with the
  reference silicon crystal data (black).}
\label{fig014}
\end{figure}

Indeed the \comsa\ plots presented in Fig~\ref{fig014} support this
assumption. Although there are still easy recognizable differences in
the \comsa\ values calculated in ROIs of different size, the
differences are not as large as before the HTHV annealing. The
values after HTHV annealing  approach the appropriate reference
silicon values.  Most striking, the averaged specific slope error
values are reduced by almost a factor two  to $\lesssim
0.1~\mu$rad/mm$^2$, becoming very close to the reference silicon
value.

It is tempting at this point to make a general conclusion that the HTHV annealing
improves substantially the Bragg-plane slope errors in diamond crystal
and makes them close to those of silicon. This is most probably true if
we are working with very high quality crystals like
VB5. Unfortunately, this conclusion is not universally applicable. The
HTHV annealing of diamond crystal VB6, which features some residual
dislocation lines in the crystal center (in contrast to VB5, which is
free of such defects), does not result in the same improvements as in
the case of VB5. The very limited amount of high-quality samples available
for our studies does not allow us to make a universal conclusion. As
stated before, most probably, the HTHV annealing is efficient in
improving Bragg-plane slope errors and pushing them to the
silicon reference limit only in high-quality diamond crystals,
that have no dislocations, no stacking faults, and so forth, in the region of
interest. Improvements are likely due to annihilation of vacancies,
impurity atoms, and so forth. These statements should be confirmed by studies
on a larger set of high-quality diamond crystals.

\section{Conclusions and outlook}

Perfect crystals with flat Bragg planes are a prerequisite for
wavefront preservation in Bragg diffraction.  We use an x-ray rocking
curve imaging technique to study the question of the
smallest-achievable Bragg-plane slope errors in the best available
synthetic diamond crystals and to compare them with those in
highest quality reference silicon crystals.

We show that the smallest specific slope errors in the best diamond
crystals are about \comss ~$\simeq 0.15-0.2~\mu$rad/mm$^2$, which are
only a factor of two larger than the $\lesssim 0.05-0.1~\mu$rad/mm$^2$ slope
errors we  measure in reference silicon crystals. 

RCI data also provides access  to the normalized specific dispersion
\fwhmssn\ of the rocking curve widths $\Delta \theta$, which is a
measure of the deviation from the largest Bragg reflectivity
achievable by perfect crystals. The best diamond crystals feature
normalized specific dispersion values \fwhmssn $\simeq$
0.01--0.013/mm$^2$ vs. $\simeq$ 0.003--0.005/mm$^2$ in silicon. These data
indicate that the local reflectivity values in the best diamond
crystals are reduced by not more than 1\% to 1.3\% from the maximum
values, in agreement with previous Bragg reflectivity studies in
diamond \cite{SSB11}.

The small slope errors are achieved not only in freestanding diamond
crystals but also in crystals firmly mounted in crystal holders,
provided the crystals are designed and machined with special
strain-relief features.

High-temperature annealing at 1450$^{\circ}$C of the best diamond
crystals may further reduce Bragg-plane slope errors to values
approaching those of silicon.

Further investigations are required to establish the
wavefront-preservation properties of the best available diamond
crystals.

\section{Acknowledgments}

Dr. Stanislav Stoupin (Cornell High Energy Synchrotron Source, CHESS)
is acknowledged for providing advanced RCI data evaluation codes for
these studies.  This material is based upon work supported by the
U.S. Department of Energy, Office of Science, Office of Basic Energy
Sciences Accelerator and Detector Research Program under Award Number
DE-SC-PRJ1006724.  Work at Argonne National Laboratory was supported
by the U.S. Department of Energy, Office of Science, Office of Basic
Energy Sciences, under contract DE-AC02- 06CH11357.


\appendix
\section{Rocking curve imaging}
\label{method}

Bragg-plane slope errors in diamond and silicon crystals are measured
using {\color{ncolor} x-ray Bragg diffraction rocking curve imaging
  (RCI), also known as sequential topography \cite{LBH99}.}

\begin{figure}
\includegraphics[width=0.5\textwidth]{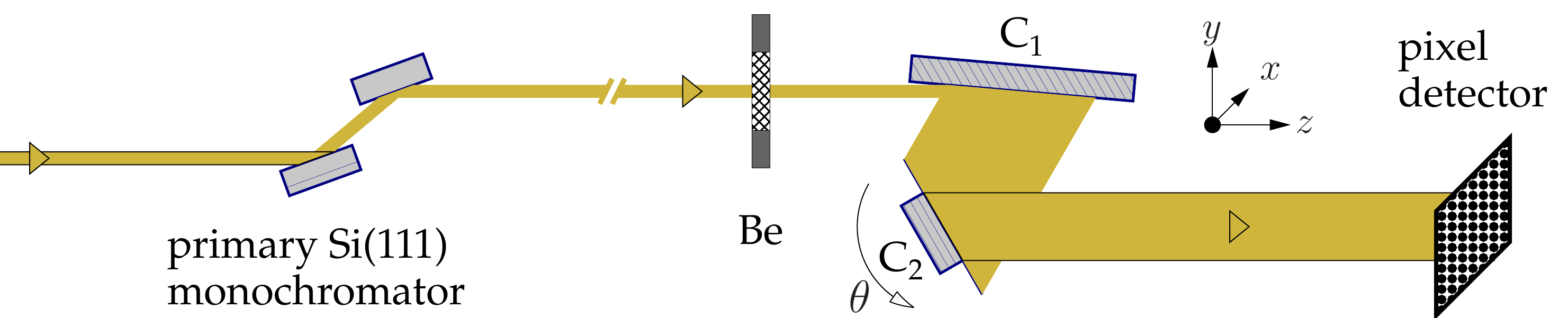}  
\caption{Layout and optical components of the rocking curve imaging
  setup at the Advanced Photon Source (APS) bending magnet beamline 1BM, comprising primary
  double-crystal Si(111) monochromator, Be windows, conditioning
  crystal C$_{\ind{1}}$, crystal under study C$_{\ind{2}}$, and pixel
  detector.}
\label{fig021}
\end{figure}

In this technique, a crystal under study is a second crystal (crystal
C$_{\ind{2}}$ in Fig.~\ref{fig021}) in a two-crystal nondispersive (or
close to nondispersive) Bragg diffraction arrangement. The first
(conditioning) crystal C$_{\ind{1}}$ is in a strongly asymmetric Bragg
reflection geometry, with the asymmetry factor $b$ chosen such that
x~rays of each photon energy have a small angular spread
$\Delta\theta_{\indrm{C1}}^{\prime}=\Delta\theta_{\indrm{C1}}^{(s)}/\sqrt{|b|}$
upon Bragg reflection (here $\Delta\theta_{\indrm{C1}}^{(s)}$ is an
angular width of the symmetric reflection), much smaller than the
Bragg reflection angular width of the second crystal
$\Delta\theta_{\indrm{C2}}\ll\Delta\theta_{\indrm{C1}}^{\prime}$. Such
asymmetric reflection from the conditioning crystal also results in an
increased cross-section of the reflected beam by a factor of $|b|$,
capable to illuminate the whole crystal C$_{\ind{2}}$ or its larger
part. If the crystals are perfect, the Bragg reflection angular
dependence (rocking curve) is very close to the intrinsic Bragg
reflection angular dependence of the second crystal under study
C$_{\ind{2}}$. The conditioning crystal is typically manufactured from
almost perfect silicon single-crystal material.

\begin{figure}
  \includegraphics[width=0.5\textwidth]{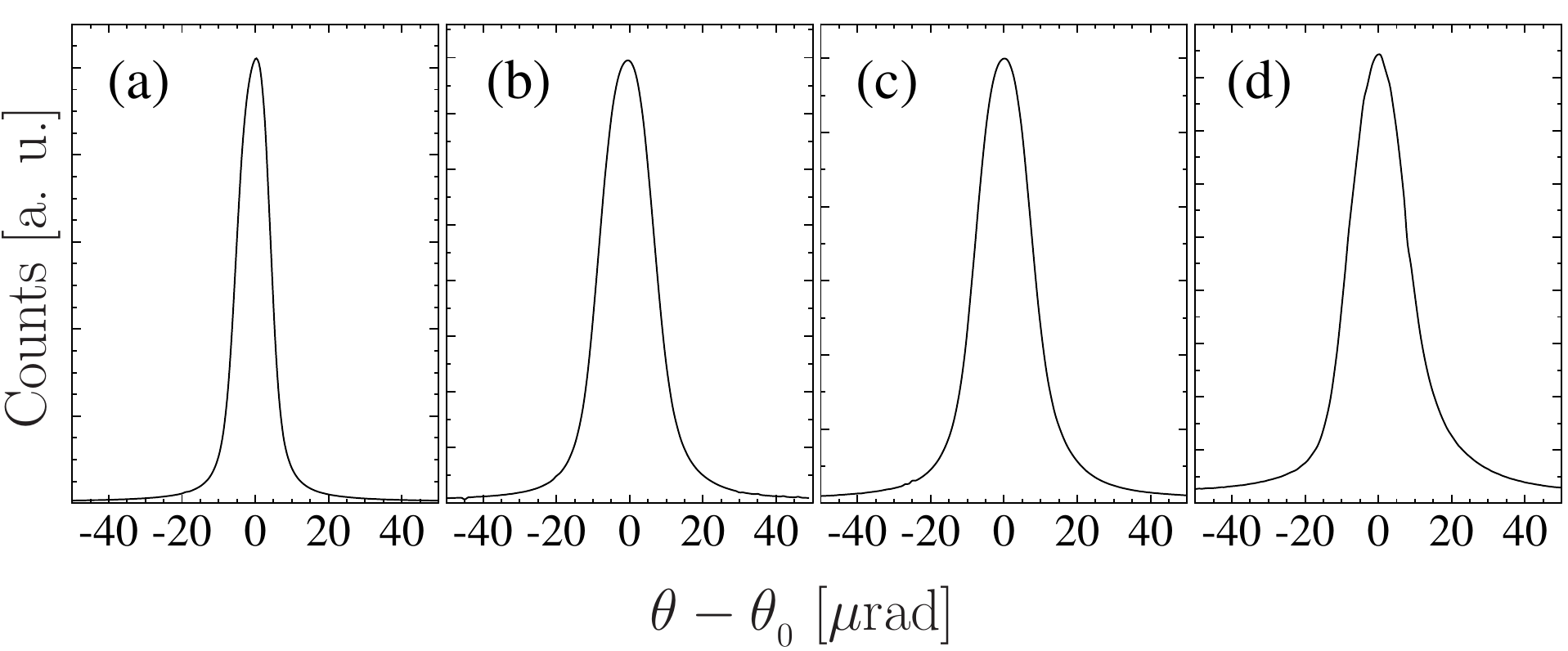}  
  \caption{Crystal-integrated angular dependences of Bragg reflectivity
  (rocking curves) of the crystals used in these experiments: (a)
  reference silicon crystal in the 531 Bragg reflection and diamond
  crystals (b) VB5, (c) VB6, and (d) D3 in the 400 Bragg
  reflection. The Bragg reflection widths (FWHM) are (a) 10.3
  $\mu$rad, (b) 17.0 $\mu$rad, (c) 17.5 $\mu$rad, (d) 18.7 $\mu$rad.
  The measurements are performed in the double crystal
  C$_{\ind{1}}$--C$_{\ind{2}}$ arrangement shown schematically in
  Fig.~\ref{fig021} with a Si PIN diode used instead of the pixel
  detector.  See text for more details.}
\label{fig022}
\end{figure}

A pixel area x-ray detector is used to measure x-ray Bragg reflection
images sequentially at different incidence angles of x-rays to the
Bragg-reflecting atomic planes of the second
crystal. Figures~\ref{fig001a}(a) and \ref{fig001}(a) show examples of
such images measured in silicon and diamond crystals,
respectively. They were recorded at the crystal-integrated Bragg
reflection rocking curve peaks. Examples of crystal-integrated rocking
curves are shown in Fig.~\ref{fig022}.  The procedure can be seen as
measuring rocking curves at particular locations of the second crystal
with the area detector pixels.  The rocking curves measured with each
detector pixel are used to calculate Bragg reflection maps.
Figures~\ref{fig001a}(b) and \ref{fig001}(b) show examples of the
color maps of the Bragg reflection angular widths $\Delta \theta$
(full width at half maximum, FWHM).  Figures~\ref{fig001a}(c) and
\ref{fig001}(c) show examples of the color map of the Bragg reflection
peak relative angular positions $\theta$, evaluated as a center of
mass (COM) of the rocking curves. The RCI maps are calculated using a
dedicated code \cite{Stoupin15}. The microscopic structure defects can
be derived from the Bragg reflection FWHM maps. The mesoscopic and
macroscopic crystal strain and Bragg planes slope errors can be best
evaluated from the COM maps.

We use an x-ray diffraction sequential topography setup \cite{SST16}
at x-ray optics testing beamline 1BM \cite{Macrander2016} at the
Advanced Photon Source (APS). The setup enables rocking curve mapping
with a submicroradian angular and 2.6-$\mu$m spatial resolution. The
latter is defined by a 6.5-$\mu$m pixel size of ANDOR Neo-5.5 sCMOS
camera and a $\times$2.5 magnification of the collecting light optic.

We study in this paper diamond crystal plates in the (100) orientation
using 400 Bragg reflection of 8-keV x-rays with Bragg's angle
$\theta_{\ind{400}}=60.345^{\circ}$.  The expected in theory Bragg
reflection width is $\Delta\theta_{\ind{400}}=14.7~\mu$rad
(FWHM). Bragg reflection 531 from silicon with a Bragg's angle
$\theta_{\ind{531}}=57.6^{\circ}$ for 8-keV x-rays
($\Delta\theta_{\ind{531}}=\Delta\theta_{\indrm{C1}}^{(s)}=9.8~\mu$rad)
is the best match for the closest-to-nondispersive double crystal
setting with the 400 Bragg reflection from diamond. This reflection is
used with the conditioning silicon crystal in the present studies. The
crystal is asymmetrically cut with the asymmetry angle
$\eta=55.6^{\circ}$, resulting in the asymmetry parameter
$b=-\sin(\theta_{\ind{531}}-\eta)/\sin(\theta_{\ind{531}}+\eta)\simeq
-1/26$.  The angular spread of monochromatic x-rays upon such Bragg
reflection is
$\Delta\theta_{\ind{531}}^{\prime}=\Delta\theta_{\ind{531}}/\sqrt{|b|}=1.9~\mu$rad,
which is much smaller than $\Delta\theta_{\ind{400}}$.

The $x$- and $y$-axis in Figs.~\ref{fig001a}(a)-(c),
~\ref{fig002}(a)-(c), as well as in other x-ray Bragg diffraction
images and RCI maps, correspond to the detector coordinates. The
diffraction plane is parallel to the $y$-axis.  The x-ray Bragg
diffraction images and RCI maps therefore appear to be contracted in
the $y$-direction by a factor of $\sin\theta_{\ind{531}}\simeq
\sin\theta_{\ind{400}}\simeq 0.87$. 

The examples presented in Fig.~\ref{fig001a} are based on the
measurements with a silicon crystal in the symmetric 531 Bragg
reflection as the second crystal under study.  Because Bragg
reflections from both crystals are the same in this case, the crystals
are in perfect nondispersive setting, with the rocking curves
unaffected by the energy and angular spread of x-rays, predominantly
determined by the Bragg reflection parameters of the second crystal.
Both silicon crystals (conditioning and reference) were manufactured
from highest quality (high-resistivity), almost perfect single-crystal
material.

The measurements with silicon C$_{\ind{2}}$ crystal are performed to
establish a reference for diamond crystals and to benchmark the
ultimate performance of the RCI setup used in this studies.  They are
addressed in more detail in Section~\ref{Si}.

\section{Moderating the impact of beamline wavefront distortions}
\label{moderate}

As x~rays propagating from the bending magnet source travel to the
crystal under study they interact with numerous beamline optical
components, including as Bragg-reflecting silicon crystals in the
primary high-heat-load monochromator, two successive beryllium
windows, and the conditioning crystal (see Fig.~\ref{fig021}).
Unfortunately, these interactions -- and primarily with the two
0.25-mm-thick Be-windows -- introduce significant wavefront
distortions. The contrast observed on the x-ray Bragg diffraction
images of the reference silicon crystal in Fig.~\ref{fig001a}(a), such
as the rippled background and other irregular solitary features are
due to the wavefront distortions and have nothing to do with the
crystal imperfections. These distortions result in local fluctuations
of the direction and of the angular distribution of the x~rays
incident on the crystal under study and therefore may perturb the
local values of the Bragg reflection widths $\Delta \theta$ and peak
positions $\theta$, and therefore falsify genuine RCI maps and RCI
characteristics. Indeed, the RCI maps in Figs.~\ref{fig001a}(b) and
(c) inherit the rippled background structure of the x-ray Bragg
diffraction image in Fig.~\ref{fig001a}(a).  This of course affects
the calculated RCI characteristic values, such as \coms and \fwhms ,
and may lead to false conclusions regarding crystal properties.

The distortions observed in Fig.~\ref{fig001a}(a) have a
characteristic length from $\simeq 10~\mu$m to $\simeq 100~\mu$m.  The
characteristic length of the Bragg-plane slope variations in the
defect-free regions are typically larger.  Therefore, the negative
impact of the wavefront distortions on the RCI maps and on the
characteristic values can be mitigated by smoothing these distortions.
A standard procedure in such cases is adding together (binning) the
signal recorded by $N\times N$ adjacent area detector pixels to
achieve better signal-to-noise ratio and to minimize the effects of
observation errors, albeit at a cost of resolution.

Figures~\ref{fig002a}(a) and \ref{fig002a}(b) show plots of RCI
characteristic values \coms\ and \fwhms, respectively, as a function
of the binning number $N$ calculated in various crystal ROIs. The blue
markers correspond to a $4\times 4$~mm$^2$ crystal ROI. The black
markers correspond to averaged RCI values calculated in four
equal-size $2\times 2$~mm$^2$ ROIs, while the red markers correspond
to averaged RCI values calculated in sixteen equal-size $1\times
1$~mm$^2$ ROIs. The error bars indicate the scattering range of the
values we are averaging.  The particular ROIs are indicated by
appropriate colors in Fig.~\ref{fig002a}(c).

The RCI characteristic values \coms\ and \fwhms\ of the reference silicon
crystal first decrees rapidly with $N$ and then reach steady
values. The binning procedure minimizes the detrimental effect of the
wavefront distortions due to the beamline optics. The genuine RCI
characteristic values \coms\ and \fwhms\ of the reference silicon are close
to those obtained with large binning numbers.

The binning procedure can ``improve'' the RCI characteristic values
\coms\ and \fwhms\ only if the wavefront contrast in x-ray Bragg
diffraction images is larger than the contrast due to crystal defects.
If a crystal is damaged or badly strained the binning procedure cannot
help.  Figures~\ref{st2-1}(a$_{\ind{1}}$)-(c$_{\ind{1}}$) show  an
example of x-ray Bragg diffraction images and RCI maps of a diamond
crystal significantly strained by laser machining.  In this case
binning cannot ``improve'' substantially the specific Bragg-plane
slope errors \comssa , shown by green markers and lines in
Fig.~\ref{st2-2}.  The improvement in Bragg-plane slope errors occurs
due to crystal annealing, as other data in Fig.~\ref{st2-2}
demonstrates.

\section{High-temperature high-vacuum annealing}
\label{HTHV}

\begin{figure}
      \includegraphics[width=0.35\textwidth]{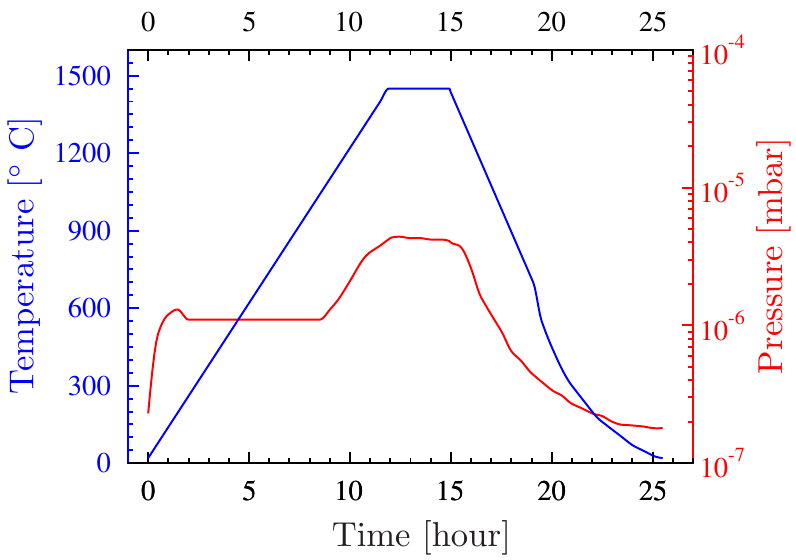}
  \caption{Time-dependence of temperature and pressure in the high-temperature high-vacuum furnace during simultaneous annealing of diamond crystals VB5 and VB6.}
\label{tp-time}
\end{figure}

A Red Devil G vacuum furnace manufactured by R.D. Webb Company
Inc. was used for high-temperature high-vacuum
annealing. Figure~\ref{tp-time} shows typical temperature and pressure
profiles in the furnace during the annealing process.  The heating
rate was 2$^{\circ}$C/min. The cooling rate was
-3$^{\circ}$C/min. From 700$^{\circ}$C the heating was switched off
and the furnace was allowed to cool naturally.

The vacuum is $\simeq 10^{-7}$~mbar at the beginning at room
temperature, but it increases to $\simeq 4\times 10^{-6}$~mbar at
1450$^{\circ}$C.  It is very important to have high-vacuum conditions
to avoid diamond damage. However, even under such high-vacuum
conditions the color of diamond changes to light gray. The subsequent
medium-temperature annealing in air erases the gray color and makes
diamond transparent again, suggesting that the color change is a
surface rather than a bulk effect.

\end{document}